\documentclass{superfri}
 \usepackage{array,multirow}
 \usepackage{amsmath}
 \usepackage{mathrsfs}
 \usepackage{fancyhdr}
 \usepackage{verbatim}
 \usepackage{color}
 \usepackage{algorithm}
 \usepackage{amsmath}
 \usepackage{caption}
 \usepackage{algorithmic}
 \usepackage{array}
 \usepackage{graphicx}

\usepackage{graphicx}
\usepackage[hidelinks]{hyperref}

\begin{document}

\author{Grzegorz Korcyl\footnote{Department of Information Technologies,  Faculty of Physics, Astronomy and Applied Computer Science, Jagiellonian University, Krak\'ow, Poland} \and  Piotr Korcyl\footnote{Institut f\"ur Theoretische Physik, Universit\"at Regensburg, Regensburg, Germany}\textsuperscript{,}\footnote{M. Smoluchowski Institute of Physics, Jagiellonian University, Krak\'ow Poland}
}

\title{Investigating the Dirac Operator Evaluation with FPGAs}

\maketitle{}

\begin{abstract}%
In recent years the computational capacity of single Field Programmable Gate Array (FPGA) devices as well as their versatility have increased significantly. Adding to that fact the High 
Level Synthesis frameworks allowing 
to program such processors in a high level language like C++, makes modern FPGA devices a serious candidate as building blocks of a general purpose High Performance Computing 
solution. In this contribution we describe benchmarks which we performed using a kernel from the Lattice QCD code, a highly compute-demanding HPC academic code for elementary particle 
simulations on the newest device from Xilinx, the U250 accelerator card. We describe the architecture of our solution and benchmark its performance on a single FPGA device 
running in two modes: using the external or embedded memory. We discuss both approaches in detail and provide estimates for the necessary memory throughput and the minimal 
amount of resources needed to deliver optimal performance depending on the available hardware. Our considerations can be used as guidelines for estimating the 
performance of larger, many-node systems.

\end{abstract}












\section*{Introduction}


Quantum Chromodynamics is the theory describing the interactions of quarks and gluons, explaining why the latter form bound states such as protons and neutrons. One of the 
characteristic features of this theory is that in the low energy regime quarks and gluons form a strongly coupled system. As a consequence it is difficult to extract predictions for the
properties of such a system from First Principles of Physics. Up to now the only available computational tool allowing for such calculations are numerical simulations (Monte Carlo simulations) 
of a discretized version of the theory, called Lattice Quantum Chromodynamics (LQCD). Traditionally, physicists working in the field of LQCD searched for the most performant, vector machines consisting of a large number of compute nodes, and have designed many new HPC solutions: QCDOC \cite{qcdoc}, APE \cite{ape}, QPACE \cite{qpace}, just to name a few. Currently GPU and ARM processors are considered for the next generation of supercomputing machines and it is an open question whether FPGA devices could be used as an alternative.

In the discretized version of Quantum Chromodynamics the basic degrees of freedom are associated 
to each point of a four-dimensional grid  representing a finite volume of four-dimensional space-time. The sizes of such volumes vary from $V = 10^6$ up to $V = 10^8$ points. The most compute intensive part of any such simulation is the inversion of the Dirac matrix, which is of the
size $(24 V) \times (24 V)$. The matrix has a sparse structure because it describes the nearest-neighbour interactions. 
The Dirac matrix $D(n,m)^{AB}_{\alpha \beta}$ acting on the vector $\psi(n)$ can be written down as follows \cite{gattringer}

\begin{multline}
D(n,m)^{AB}_{\alpha \beta} \psi_{\beta}^B(m) = (m_q + 4) \psi_{\alpha}^A(n) +\\+
 \frac{1}{2} \sum_{\mu=0}^3 \Big[ U_{\mu}^{AB}(n) P^{-\mu}_{\alpha \beta} \psi_{\beta}^B(n+\hat{\mu}) +
U^{\dagger, AB}(n-\hat{\mu}) P^{+\mu}_{\alpha \beta} \psi_{\beta}^B(n-\hat{\mu}) \Big]
\label{eq. dirac op}
\end{multline}

The most elementary computational block is the evaluation of the single stencil, i.e. evaluation of the right hand side of \eqref{eq. dirac op} for a given value of the index $n$. Note, that
the coefficients of the $D(n,m)^{AB}_{\alpha \beta}$ matrix differ for each $m$, i.e. the $U$ complex-valued $3 \times 3$ matrices and $\psi$ complex-valued 3-element vectors depend on the 
position $m$. Therefore, each stencil involves loading of eight $U(n)$ matrices and nine spinor fields from the neighboring lattice sites, which in total corresponds to 360 input words. In the case of double precision this amounts to 2880 input bytes. One can exploit the structure of the $SU(3)$ matrices and parametrize them in terms of 10 input words each, instead of 18 in the naive formulation (9 real and 9 imaginary entries). We return to this point in Section \ref{sec. stream}. 
The $U \times \psi$ matrix-vector multiplications require 1464 floating point operations for complex additions and multiplications. $P^{\pm}$ are real-valued $4\times4$ constant 
matrices, $m_q$ is a real parameter corresponding to the quark mass, $\mu$ labels directions in the four-dimensional space-time. Repeated indices are summed within the ranges: 
$\alpha,\beta=1,\dots,4$, $A,B=1,2,3$. For unexplained notation please see \cite{gattringer} or \cite{pg}.
One of the simplest algorithms allowing to invert such a matrix is an iterative conjugate gradient algorithm. The relevance 
of this algorithm is demonstrated by the fact that the HPCG benchmark was introduced since November 2017 as a new ranking of supercomputers published by the TOP500 organization.
%
%
%
Such benchmark differs from the traditionally used Linpack benchmark where the employed matrix was dense. The argument behind the HPCG benchmark is that in many cases sparse matrix computations are
more representative of the variety of HPC applications which run on a supercomputer.  
Indeed, the iterative solver of the type of conjugate gradient is, for instance, at the heart of Monte Carlo simulation of QCD. 

The rest of this article is organized as follows. In the next section we specify the details of the implemented algorithm 
as well as summarize the description of the kernel which is being hardware accelerated. Subsequently in the following section 
we propose two implementations on the FPGA devices which differ by the location where the main data is stored, either these 
are registers in the programmable logic, or an external DDR memory bank attached to the programmable logic. In section \ref{sec. discussion} we compare and discuss the achieved performances using both approaches. Eventually we conclude and point to future research directions.

\section{Kernel Description}
\label{sec. single}

In this work we consider an improved version of the conjugate gradient algorithm which allows us to test different floating 
and fixed point precision without a deterioration of the ultimate solution. Similar considerations for GPU were presented in \cite{gpu}. The algorithm intertwines iterations in low and 
high precision, working mainly in low precision and correcting a possible systematic error by a high precision iteration. 
Our algorithm follows the one suggested in \cite{strzodka} and is shown in Algorithm \ref{algorithm}. 
We provide an exact form of the mixed precision conjugate gradient algorithm implemented in this work to show which parts have been hardware accelerated and what is the interplay between parts of the algorithm requiring implementations in different precision. In both cases the most time consuming part are matrix 
multiplications in lines \ref{dd high 2}, \ref{dd low} and \ref{dd high} of Algorithm \ref{algorithm}. 

\begin{algorithm}
\caption{Residual Guided CG algorithm}
\begin{algorithmic}[1]
\label{algorithm}
\STATE $\psi^{\textrm{high}} \gets \psi^{\textrm{high}}_0$
	\STATE $r_0^{\textrm{high}} \gets \eta^{\textrm{high}} - \big( D^{\dagger}D \big)^{\textrm{high}} \psi^{\textrm{high}}$ \label{dd high 2}
\STATE $s_0^{\textrm{high}} \gets ||r_0^{\textrm{high}}||$
\STATE $r_0 \gets \frac{r_0^{\textrm{high}}}{s_0^{\textrm{high}}}$
\STATE $l \gets 0$
\WHILE {$s^{\textrm{high}} \geq r^{\textrm{high}}_{min}$}
    \STATE $n \gets 0$
    \STATE $\psi_0 \gets 0$
    \STATE $r_0 \gets \frac{r_{l+1}^{\textrm{high}}}{s_{l+1}^{\textrm{high}}}$
    \STATE $p_0 \gets p_k - \big( r_0 \cdot p_k \big) r_0$
    \STATE $\alpha_0 \gets 0$
    \STATE $\beta_0 \gets \frac{s_{l+1}^{\textrm{high}}}{s_l^{\textrm{high}} \rho_k}$
    \WHILE {$n < k$} 
	\STATE $q_n \gets D^{\dagger} D p_n$ \label{dd low}
        \STATE $\alpha_n \gets \frac{\rho_n}{p_n \cdot q_n}$
        \STATE $\psi_{n+1} \gets \psi_n + \alpha_n p_n$
        \STATE $r_{n+1} \gets r_n - \alpha_n q_n$
        \STATE $\rho_{n+1} \gets r_{n+1} \cdot r_{n+1}$
	    \STATE $\beta_n \gets \frac{\rho_{n+1}}{\rho_n}$
	    \STATE $p_{n+1} \gets r_{n+1} + \beta_n p_n$
	    \STATE $n \gets n+1$
    \ENDWHILE
    \STATE $\psi^{\textrm{high}}_{l+1} \gets \psi_l^{\textrm{high}} + s_l^{\textrm{high}}\big( \psi_k + \alpha_k p_k \big)$
	\STATE $r_{l+1}^{\textrm{high}} \gets b^{\textrm{high}} - \big(D^{\dagger}D\big)^{\textrm{high}} \psi_{l+1}^{\textrm{high}}$ \label{dd high}
    \STATE $s_{l+1}^{\textrm{high}} \gets || r_{l+1}^{\textrm{high}}||$
    \STATE $l \gets l+1$
\ENDWHILE
\label{alg. cg}
\end{algorithmic}
\end{algorithm}
We wish to hardware accelerate them 
and briefly summarize the FPGA 
implementation of these kernel functions. We follow what was presented in \cite{pg}. In particular that Reference contains 
a description of C++ data structures used for the implementation as well as relevant details of the memory allocation which allows for
a fully pipelined execution of the kernel. Fragments of C++ and HLS directives codes are provided and discussed in that Reference. 

For both, high and low, precisions of the kernel the implementation 
is similar: a single function involves a loop over a subvolume and an evaluation of the 
stencil for each site of the lattice. The evaluation of a single stencil is fully parallelized as far as the data dependencies allow and all stencils are pipelined.

\begin{figure}
\centering
\includegraphics[width=320pt]{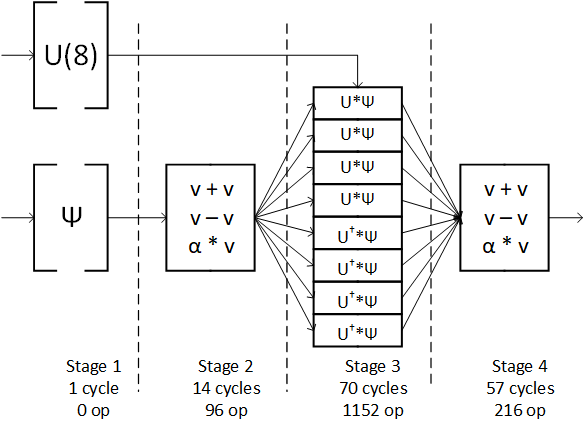}
\caption{Computation sequence of the stencil solver}
\label{fig. kernel}
\end{figure}

All operations involved in the estimation of a single stencil are graphically shown in Fig.~ \ref{fig. kernel}. The evaluation naturally splits into 4 stages. The clock cycles provide an estimate of the amount of parallelization and correspond to the number of clock cycles required to finish the computation at a given stage in double precision. In the first stage all the necessary data is copied from the BRAM memory blocks to local registers which requires only one clock cycle. In stage 2 linear combinations of input data, 8 additions and 8 subtractions of vector type, are evaluated. They are all performed in parallel, taking 14 clock cycles, which corresponds to a single addition of double numbers in programmable logic. The most 
compute intensive stage 3 involves $SU(3)$ matrix by vector multiplications. In total 1152 operations are performed. Complete parallelization allows to execute them in a 5-layer operation cascade taking in total $5*14=70$ cycles. Finally, at stage 4 all contributions are added up to the final result. Because of the dependencies between consecutive partial results this creates a 4-layer operation cascade, which in total takes $57=(4*14)+1$ clock cycles, 4 additions plus one data copy. Overall, the kernel requires 142 clock cycles and a total of 1464 basic operations to compute the final result since the reception of the input data. The kernel is fully pipelined: i.e. it can accept new input data at each clock cycle and produce the results with latency of 142 cycles. 

\section{Two Approaches}

There are two approaches one can follow in order to provide to the kernel the required data. One can divide the entire problem into small parts such that the entire set of data for a single part fits into the BRAM memory of the device. Alternatively, one can store the entire set of data in the DDR die attached to the programmable logic and stream the data through the link. We discuss below the performances of both solutions. 

\subsection{Smaller Lattice Stored in BRAM Memory}
\begin{figure}
\centering
\includegraphics[width=320pt]{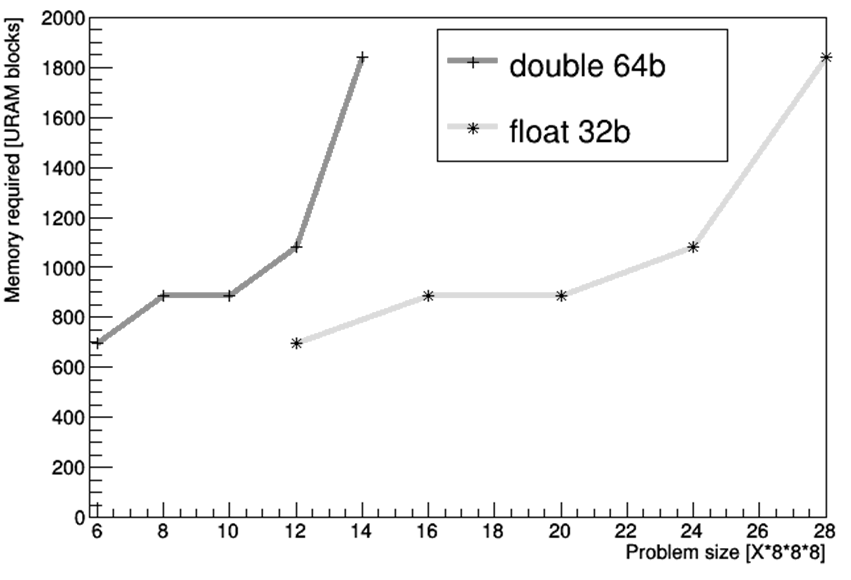}
\caption{Memory usage as a function of the initiation interval}
\label{fig. memory}
\end{figure}

This is the approach we followed in \cite{pg}. We showed that lattices up to the size of $12 \times 8^3$ data points in each direction in double precision can fit into the internal memory of the programmable logic of the FPGA devices available currently on the market. In Fig.~ \ref{fig. memory} we show the required number of URAM blocks for a given size of the lattice for single and double precision. As can be seen in that figure the storage requirements are not linear because in order to allow the compiler to take advantage of the natural parallelism of FPGA devices it is crucial to store data in PL in as many separate PL local registers blocks as possible. This is due to the fact that in a single PL clock cycle only one memory element can be read from the BRAM block. In the computation of a single stencil one needs eight different $U$ matrices and we impose that they are stored separately. Although this requires duplicating the amount of stored data, the matrices $U(n)$ and $U^{\dagger}(n)$ are stored separately, the gain is considerable. The HLS directives ensuring such memory allocation were described in \cite{pg}. Thanks to that the stencil evaluation can be fully pipelined, i.e. the hardware block can accept new input data at each clock cycle. The  resulting performance simulated in software is 812 GFLOPs for single precision and 406 GFLOPs for double precision with the PL running at 300 MHz.

\subsection{Larger Lattice Streamed from the DDR Memory}
\label{sec. stream}
\begin{figure}
\centering
\includegraphics[width=320pt]{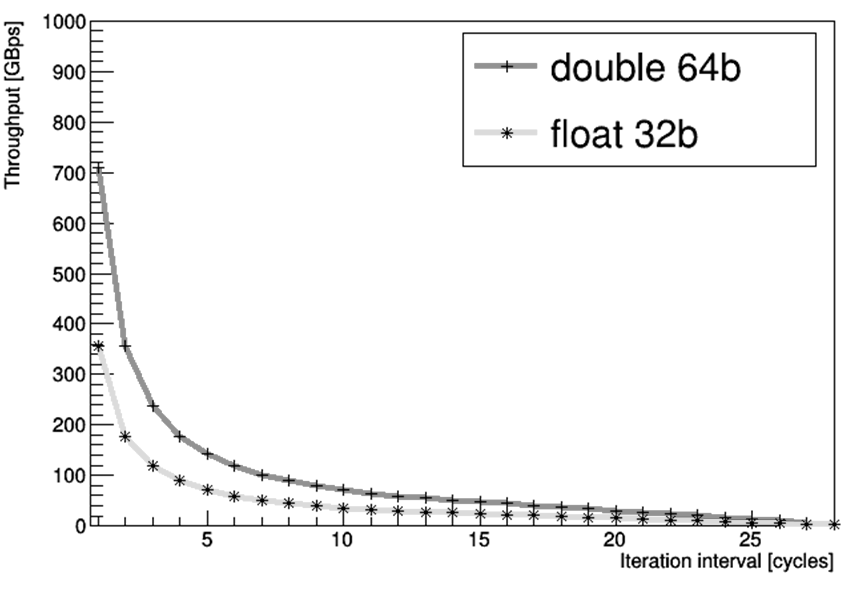}
\caption{Transmission rates as a function of the initiation interval}
\label{fig. rates}
\end{figure}
\begin{figure}
\centering
\includegraphics[width=320pt]{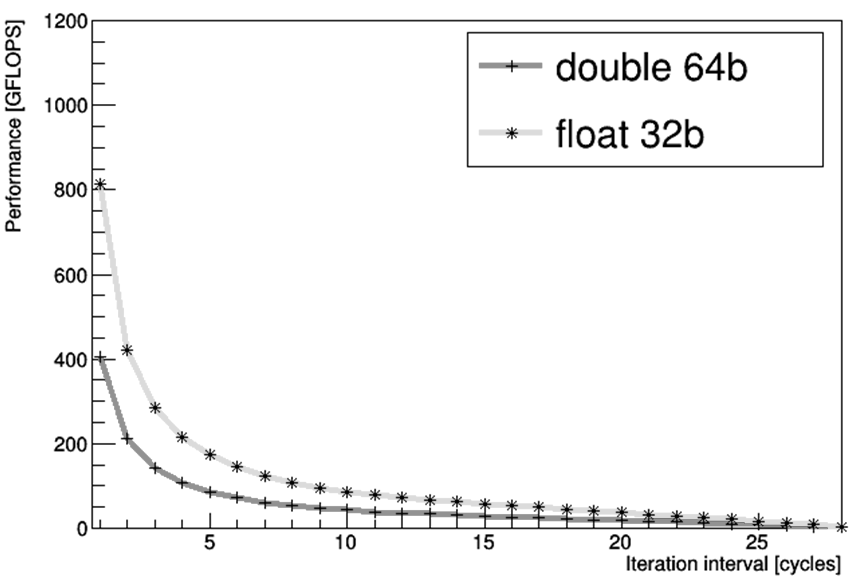}
\caption{Performance as a function of the initiation interval}
\label{fig. peformance2}
\end{figure}
The way to operate on larger data sets is to keep the data in the DDR die attached to the programmable logic and process the data in a streaming mode. This was investigated on the Maxeler system in \cite{janson}. 
The $U$ matrices and $\psi$ spinors are prepared beforehand into sets corresponding to consecutive stencils and are streamed continuously from the DDR into the logic. The limitation of this solution is the throughput of the memory link between the DDR and the logic. Using SDAccel and an openCL implementation of the CG algorithm we verified that for the Xilinx U250 device one can send 256B from the DDR memory to the PL part per clock cycle working at the frequency of 300 MHz. Four channels are available aggregating to 77 GBps throughput. In order to decrease the amount of data transferred we change the representation of the $U$ matrices and following \cite{rainer} we use a 10 parameter parametrization. We trade two more parameters and avoid computing trigonometric functions in the programmable logic. The reduced set of data translates to an initiation interval of 5 and 9 clock cycles for the compute kernel for single and double precision respectively, i.e the programmable logic has to wait 5/9 clock cycles to gather enough data to start a new computation. The performance in that case would approximately be equal to 86 and 46 GFLOPs respectively, which is comparable to the one quoted in \cite{janson} on the Maxeler system. However, if we also count the additional operations needed to recover the $U$ matrices from their reduced form, the achieved sustained performance reaches 194 GFLOPs for single precision. In Fig.~ \ref{fig. rates} we show how the required throughput depends on the initiation interval. The calculation assumes the reduced form of the $U$ matrices. The smaller the initiation interval is, the shorter is the time in which the data has to be transferred. We show the throughput estimates for single and double floating point precision. Knowing the throughput between the DDR and the programmable logic on a given device one can easily read the corresponding minimal initiation interval and henceforth the resulting performance, which is shown in Fig.~ \ref{fig. peformance2} for both the single and double precision case.
\begin{figure}
\centering
\includegraphics[width=320pt]{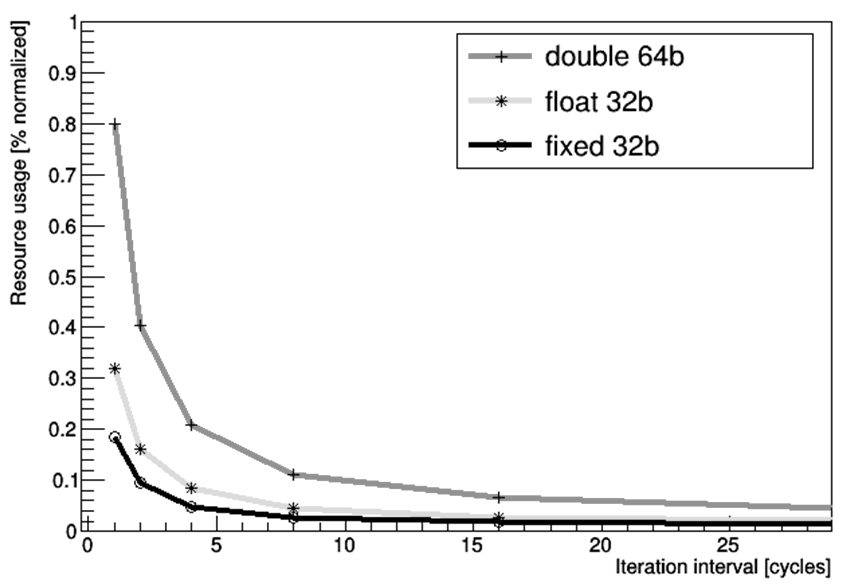}
\caption{Resources consumption as a function of the initiation interval}
\label{fig. resources}
\end{figure}

Finally, in Fig.~ \ref{fig. resources} we show how the hardware resource consumption depends on the initiation interval for single and double floating point precision but also for a more FPGA friendly 32 bit fixed point data format. In this streaming scenario one can relax the initiation interval of one clock cycle imposed in the first approach. The memory throughput being the bottleneck, one can implement the kernel with a lower initiation interval because in any case several clock cycles are needed to collect all the necessary data for a single stencil computation. The figure shows an indicative percentage of all available resources counting together all DSP, LUTs and BRAM blocks. We see that in the described case where the memory throughput imposes an initiation interval of 5 clock cycles the compute kernel uses only 20\% of the available resources for double precision.

\section{Discussion}
\label{sec. discussion}

The presented results allow to understand various constraints limiting the performance of the investigated kernel on FPGA devices. Starting from the embedded memory scenario, the practical problems that are being analyzed are  larger by a factor of the order of 4096. One would probably use that amount of FPGA devices running in parallel and exchanging boundary data directly from and to the programmable logic through the embedded transceivers. On the other hand, in the external memory scenario in principle the entire set of data could be stored in the DDR. However, the wall clock time to the solution on a single FPGA device would be impractically long. In that case one would also resort to a many-node system where the computations could be speed up by running them in parallel. In principle neither of the two scenarios is obviously superior. The number of required nodes can be different in both solutions and the details would depend essentially on the memory throughput of the FPGA device used. With the numbers provided above such estimations can be put on a solid ground.

\section*{Conclusions}

In this contribution we discussed the applicability of FPGA devices to High Performance Computing solutions. As a benchmark we used the academic code for Monte Carlo simulations 
of Quantum Chromodynamics. In traditional computer architectures this code is memory bound due to the unfavorable ratio of the amount of data to be loaded to the amount of floating point operations
to be executed be the most elementary kernel function.  On
the available programmable logic hardware the problem turns out to be memory bound in the scenario where the data is streamed from the DDR die, which will be considerably improved with the arrival of Xilinx Alveo U280 cards with a 480 GB/s memory bandwidth between DDR and programmable logic. In the scenario where the data is stored in the embedded memory, the problem's limitation is the available size of the internal memory. Both cases seem to be scalable and thus offer a viable proposal for a larger scale infrastructure.

\section*{Acknowledgments}

This work was in part supported by Deutsche Forschungsgemeinschaft under Grant No. SFB/TRR 55
and by the polish NCN grant No. UMO-2016/21/B/ ST2/01492, by the Foundation for Polish Science grant no. TEAM/2017-4/39 and by the Polish Ministry for Science and Higher Education grant no. 7150/E-338/M/2018.
The project could be realized thanks to the support from Xilinx University Program and their donations. P.K. acknowledges support from the NAWA Bekker fellowship and thanks Universit\`{a} degli Studi di Roma Tor Vergata for hospitality during which this work was finalized. \\

\emph{This paper is distributed under the terms of the Creative Commons Attribution-Non Commercial 3.0 License which permits non-commercial use, reproduction and distribution of the work without further permission provided the original work is properly cited.}

\bibliography{references}

\end{document}